# Simple and Efficient Contract Signing Protocol

Abdullah M. Alaraj
Information Technology Department
College of Computer, Qassim University
Saudi Arabia

*Abstract*—In this paper, a new contract signing protocol is proposed based on the RSA signature scheme. The protocol will allow two parties to sign the same contract and then exchange their digital signatures. The protocol ensures fairness in that it offers parties greater security: either both parties receive each other's signatures or neither does. The protocol is based on offline Trusted Third Party (TTP) that will be brought into play only if one party fails to sign the contract. Otherwise, the TTP remains inactive. The protocol consists of only three messages that are exchanged between the two parties.

## I. INTRODUCTION

Contracts play an important role in many business transactions. Traditionally, paper-based contracts are signed by the transacting parties who need to be present at the same venue and at the same time. Each party signs a copy of the contract for every contracting party so that every party has a copy of the signed contract.

If the parties, however, are not able to meet to sign the paper-based contract, then signing an electronic contract is an alternative. The problem with signing electronic contracts, however, is exchanging the signatures of the parties, especially where there is a lack of trust between parties. One party may send the other party their signature on the contract but may not receive the signature of the other party in return. To solve the problems of exchanging digital signatures, contract signing protocols are used [3, 4, 5, 9, 10]. Contract Signing Protocols ensure that either contracting parties receive each other's signature or none does.

In this paper, a new, efficient contract signing protocol is proposed. The proposed protocol is based on offline trusted third party (TTP) that brought into play only if one party fails to send their signature on the contract. In the normal execution of the protocol, the two parties will exchange their signatures directly.

This paper is organized as follows. Related work is presented in section II. Section III presents the proposed protocol that comprises the exchange protocol and dispute resolution protocol. The analysis of the proposed protocol is discussed in section IV. The comparison of the proposed protocol with related protocols is presented in section V.

## II. RELATED WORK

Early contract signing protocols (as in [7, 16]) allow the parties to exchange their signatures directly without any involvement from third party. That is, the parties gradually exchange their signatures in part until both signatures are complete. If one party fails to send an additional part of the signature, the other party works to search for that remaining part. The gradual exchange protocols are based on the assumption that the two parties have the same computational power to ensure fairness. However, in most applications this assumption is not realistic [5]. The gradual exchange protocols require a large number of rounds to complete the exchange of signatures.

To overcome the problems of gradual exchange of signatures, a trusted third party (TTP) is used in contract signing protocols. The TTP helps the contracting parties to exchange their signatures in a reliable and secure manner. The TTP can be used online or offline.

In the online-based third party contract signing protocols [as in 6, 8,10] the TTP will be actively involved in the exchange of the signatures between the parties. The parties will sign the contract and send their signatures to the TTP who will verify the signatures and if they are correctly verified the TTP will forward the signatures to the parties. The main problem with this approach is that the TTP is involved in every exchange and this may create a bottleneck. In addition to this, the fees of the third party make this a costly approach.

In the offline-based third party contract signing protocols [as in 3, 4, 5, 11, 13 (also called optimistic – 11)], the parties will directly exchange each other's signatures on a contract. If one party fails to submit their signature, the third party will be brought in to resolve any dispute. In the offline-based third party contract signing protocols, the TTP is rarely involved which reduces the cost of running TTP. Also, the turnaround time is eliminated since the parties exchange their signatures directly.

A category of offline TTP-based contract signing protocols has been proposed [3, 4, 5]. This category overcomes the farness problem by using verifiable and recoverable encrypted signatures. This approach will generally work as described below. Let's say that two contracting parties, Alice and Bob, want to exchange their signatures on a contract.

Alice will sign the contract, encrypt the signature and then send the encrypted signature to Bob. Bob will then verify the encrypted signature and if it is correctly verified, send his signature to Alice. If Alice finds that Bob's signature is correct then she will send the decryption key to Bob to decrypt her encrypted signature. If Alice fails to send the decryption key, Bob will contact the TTP to recover the decryption key.

Nenadic, Zhang and Barton[3] proposed a fair signature exchange protocol. The protocol is based on the verifiable and recoverable encryption of signatures on a contract. Alice will send her partially encrypted signature to Bob who will be able to verify it. If the encrypted signature is correctly verified then Bob will send Alice his signature. On receiving Bob's signature, Alice will verify it and if it is correctly verified then Alice will send the decryption key to Bob to decrypt the encrypted signature. If Alice does not send the decryption key, Bob will contact the TTP to recover Alice's signature.

Ateniese [4] also proposed a fair contract signing protocol. Ateniese's protocol is based on the verifiable and recoverable encryption of a signature. If Alice and Bob want to exchange their signatures on a contract then the protocol will work as follows. Alice will first sign the contract, then encrypt the signed contract with the public key of the trusted third party (TTP). Alice will then send Bob: (1) the encrypted signature, (2) evidence stating that Alice has correctly encrypted her signature on the contract. On receiving Alice's message, Bob will verify the evidence. If the evidence is valid then Bob will send his signature on the contract to Alice. On receiving Bob's signature, Alice will verify it and if it is valid then Alice will send her signature on the contract to Bob. If Alice does not send her signature to Bob or Alice's signature is invalid then Bob can contact the TTP to resolve the dispute.

Wang [5] proposed a protocol for signing contracts online. Their protocol is based on the RSA signature. If Alice and Bob are planning to exchange their signatures on a contract using Wang's protocol [5] then Alice will first split her private key into two parts d1 and d2. Only d2 will be sent to TTP. Alice will send Bob her partial signature that was signed using d1. On receiving Alice's partial signature, Bob will initiate an interactive zero-knowledge protocol with Alice to check whether Alice's partial signature is correct. If it is correctly verified then Bob will send his signature to Alice. After Alice receives Bob's signature, Alice will verify it and if it is correctly verified then Alice will send Bob the second part of her signature. If, however, Alice did not send the second part of the signature, Bob can contact the TTP to resolve the dispute.

In this paper, we propose a new approach that uses verifiable and recoverable encryption of signatures that will allow the party who receives the encrypted signature to verify it. If he / she correctly verifies the encrypted signature, then it is safe for this party to release his / her signature to the other party because the TTP can be contacted to recover the signature if the other party fails to submit his / her signature. The proposed protocol does not use the interactive zero-knowledge proofs for verifying the encrypted signature as in [4 & 5]. Rather, the contract certificate that is introduced in this paper will allow the party who receives the encrypted signature to verify it.

III. THE PROPOSED CONTRACT SIGNING PROTOCOL

A. Notations

The following represents the notations used in the proposed protocol:

- $P_a$, $P_b$, and $P_t$: parties a, b, and TTP, respectively.

- C: The contract to be signed by $P_a$ and $P_b$

- $C_{\cdot at}$: the certificate for the shared public key between $P_a$ and $P_t$. $C_{\cdot at}$ is issued by $P_t$. A standard X.509 certificate [12] can be used to implement $C_{\cdot at}$

- $Pk_x = (e_x, n_x)$: RSA Public Key [14] of the party x, where $n_x$ is a public RSA modulus and $e_x$ is a public exponent

- $Sk_x = (d_x, n_x)$: RSA Private Key [14] of the party x, where $n_x$ is a public RSA modulus and $d_x$ is a private exponent

- h(M): a strong-collision-resistant one-way hash function

- $enc.pk_x(M)$: an RSA [14] encryption of message M using the public key $pk_x$ ($e_x$, $n_x$). The encryption of M is computed as follows: $enc.pk_x(M) = M^{ex} \mod n_x$

- $enc.sk_x(Z)$: an RSA [14] decryption of Z using the private key $sk_x$ ($d_x$, $n_x$). The decryption of Z is computed as follows: $enc.sk_x(Z) = Z^{dx} \mod n_x$

- $Sig._x(M)$: the RSA digital signature [14] of the party *x* on M. The digital signature of party *x* on M is computed by encrypting the hash value of M using the private key $sk_x(d_x, n_x)$.

- C-Cert: the contract certificate. C-Cert is issued by CA. The contents of C-Cert are:
    - heSig: the hash value of the signature of $P_a$ on the contract encrypted with $pk_{at}$ i.e. "$h(enc.pk_{at}(Sig._a(C)))$"
    - hC: hash value of the contract
    - CA's signature on C-Cert

- $P_x \rightarrow P_y$: M, means party x sends message M to party y

- X + Y: concatenation of X and Y

B. Assumptions

The following represents the assumptions used in the proposed protocol:

- Channels between $P_a$, $P_b$ and $P_t$ are resilient i.e. all sent messages will be received by their intended recipients

- Parties will use the same hashing, encryption, decryption algorithms.

- $P_t$ is trusted by all parties and will not collude with any other party

- Parties $P_a$ and $P_b$ will agree on the contract before the protocol starts

- Parties ($P_a$, $P_b$ and $P_t$) already have their public keys and they are certified from CA

*C. Registration*

In the registration phase, Pa needs to do the following:

- $P_a$ will request from $P_t$ to share an RSA public key with it. The shared public key is denoted as $pk_{at} = (e_{at}, n_{at})$ and its corresponding private key is denoted as $sk_{at} = (d_{at}, n_{at})$. $P_t$ will certify the shared public key and issue the shared public key certificate $C_{\cdot at}$

- Pa will sign the contract "C" using its private key $sk_a$ as $Sig_{\cdot a}(C)$ and then send the following to CA to certify the encrypted signature and issue C-Cert:

$Sig_{\cdot a}(C) + C + C_{\cdot at}$

On receiving $P_a$'s request, CA will verify if the received signature is for the contract C included in $P_a$'s message. If so, then CA will encrypt $Sig_{\cdot a}(C)$ using the shared public key $pk_{at}$ that is included in C.at. That is, CA will compute:

enc.$pk_{at}(Sig_{\cdot a}(C))$

Then, CA will issue C-Cert that includes the items mentioned in the "Notations" section.

*D. Exchange Protocol*

The exchange protocol represents the normal execution of the protocol. It consists of the following three steps (see Fig. 1):

1. [E-M1]: $P_a \rightarrow P_b$: C + $C_{\cdot at}$ + C-Cert + enc.$pk_{at}(Sig_{\cdot a}(C))$
2. [E-M2]: $P_b \rightarrow P_a$: $Sig_{\cdot b}(C)$
3. [E-M3]: $P_a \rightarrow P_b$: $Sig_{\cdot a}(C)$

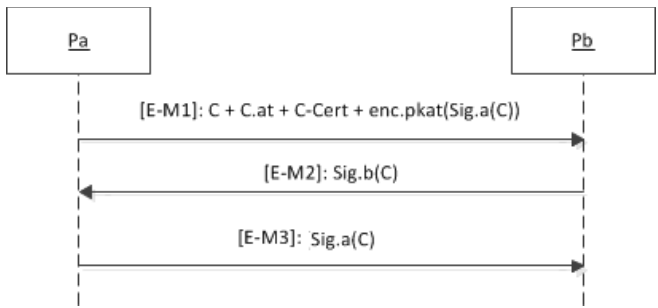

Figure 1. Exchange Protocol

Step [E-M1]: $P_a$ encrypts the signed contract with the shared public key pkat. Pa then sends the items C, C.at, C-Cert, enc.pkat($Sig_{\cdot a}(C)$) to $P_b$.

Step [E-M2]: once $P_b$ receives E-M1 then they will do the following verifications:

1. $P_b$ will verify the correctness of both $C_{\cdot at}$ and C-Cert by verifying the signatures on these certificates.
2. If the certificates are correctly verified then $P_b$ will compute the hash value of the contract and then compare it with "hC" that is included in C-Cert.
3. $P_b$ will also need to verify the correctness of the encrypted signature of $P_a$ on the contract i.e. $P_b$ will verify "enc.$pk_{at}(Sig_{\cdot a}(C))$". To verify the encrypted signature, $P_b$ will compute the hash value of "enc.$pk_{at}(Sig_{\cdot a}(C))$" then compare it with "heSig" that is included in C-Cert. If they match, it means that $P_a$ encrypted the correct signature.

If all verifications are correct then $P_b$ will sign the contract using their private key skb then will send the signed contract "$Sig_{\cdot b}(C)$" to $P_a$.

Step [E-M3]: once $P_a$ receives Sig.b (C), Pa will verify $P_b$'s signature. That is, $P_a$ will decrypt the signature to get the hash value of the contract then compare it with "hC" that is included in C-Cert. If $P_b$'s signature is correctly verified then $P_a$ will send their signature Sig.a(C) to $P_b$

Once $P_b$ receives $Sig_{\cdot a}(C)$ then $P_b$ will verify it by decrypting the signature to get the hash value of the contract and compare it with "hC" that is included in C-Cert. If the verification is correct then the received signature is correct.

Now, both $P_a$ and $P_b$ have each other's signatures on the contract. Therefore, fairness is ensured. If $P_a$ did not send E-M3 or sent incorrect E-M3 then $P_b$ can contact $P_t$ using the dispute resolution protocol to resolve the dispute.

*E. Dispute Resolution Protocol*

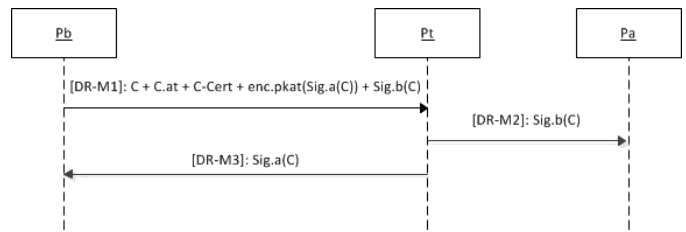

Figure 2. Dispute Resolution Protocol

If $P_b$ did not receive the step E-M3 or received an incorrect E-M3, $P_b$ can contact $P_t$ to resolve the dispute. The dispute resolution protocol consists of the following three steps (see Fig. 2):

1. [DR-M1]: $P_b \rightarrow P_t$: C + $C_{\cdot at}$ + C-Cert + enc.$pk_{at}(Sig_{\cdot a}(C))$ + $Sig_{\cdot b}(C)$
2. [DR-M2]: $P_t \rightarrow P_a$: $Sig_{\cdot b}(C)$
3. [DR-M3]: $P_t \rightarrow P_b$: $Sig_{\cdot a}(C)$

Step [DR-M1]: if $P_b$ did not receive the correct signature or did not receive the signature at all then $P_b$ will send message DR-M1 to $P_t$ to request a resolution.

Step [DR-M2]: once $P_t$ receives DR-M1 then they will do the following verifications:

- $P_t$ will verify the correctness of $C_{\cdot at}$ and C-Cert by checking the signatures on these certificates.

- If the certificates are correctly verified then $P_t$ will verify the correctness of the encrypted signature of $P_a$ on the contract i.e. enc.$pk_{at}(Sig_{\cdot a}(C))$. To verify the encrypted signature, $P_t$ will either (i) compute the hash value of enc.$pk_{at}(Sig_{\cdot a}(C))$ then compare it with "heSig" that is included in C-Cert. If they match it means that $P_a$ encrypted the correct signature, or (ii) $P_t$ has the private key "$sk_{at}$" corresponding to the shared public key so it can decrypt the encrypted signature i.e. enc.$pk_{at}$ (Sig.$_a$ (C)) and then decrypt the signature with $pk_a$ and compare the decrypted hash with "hC" that is included in C-Cert.

- $P_t$ will also verify Sig.$_b$ (C) by decrypting the signature with $pk_b$ then comparing the decrypted hash with "hC" that is included in C-Cert.

If all verifications are correct then $P_t$ will send the message DR-M2 to $P_a$ and DR-M3 to $P_b$. DR-M2 includes the signature of $P_b$ on the contract.

The signature of $P_b$ on the contract is sent to $P_a$ to ensure fairness in the case where $P_b$ contacted $P_t$ after receiving E-M1 i.e. $P_b$ may cheat by contacting $P_t$ before sending E-M2 to $P_a$.

Step [DR-M3]: $P_t$ will send Sig.$_a$ (C) to $P_b$ in DR-M3

Now, both $P_a$ and $P_b$ have each other's signature on the contract. Fairness is ensured either in the exchange protocol or in the dispute resolution protocol if $P_a$ acts dishonestly.

## IV. ANALYSIS

The fairness property in our protocol will be evaluated by studying the following four cases: (1) the first case where $P_a$ is honest and $P_b$ is dishonest, (2) the second case where $P_a$ is dishonest and $P_b$ is honest, (3) the third case where both $P_a$ and $P_b$ are dishonest, and (4) the forth case where both $P_a$ and $P_b$ are honest.

- Case 1: If $P_a$ is honest and $P_b$ is dishonest. $P_b$ acts dishonestly by sending an incorrect signature to $P_a$ or by contacting $P_t$ before sending his signature to $P_a$. In the first scenario where $P_b$ sends an incorrect signature to $P_a$, $P_a$ will check $P_b$'s signature. Then if it is incorrect, $P_a$ will not send his signature to $P_b$ in E-M3. In the second scenario where $P_b$ contacted $P_t$ before sending his signature to $P_a$, $P_t$ will check $P_b$'s request and if it is correctly verified then $P_t$ will send the resolution to both $P_a$ and $P_b$. Therefore, fairness is ensured

- Case 2: $P_a$ is dishonest and $P_b$ is honest. $P_a$ can act dishonestly by sending the incorrect E-M1, sending the incorrect E-M3 or not sending the E-M3 at all. In the scenario where $P_a$ sends incorrect E-M1, $P_b$ will verify E-M1 as described in section III. If $P_b$ finds that E-M1 is incorrect, they will not send their signature to $P_a$ in E-M2. In this scenario no one reveals their signature at this stage. In the scenarios where $P_a$ sends incorrect E-M3 to $P_b$ or $P_a$ does not send E-M3, $P_b$ can contact $P_t$ to recover $P_a$'s signature.

- Case 3: both $P_a$ and $P_b$ are dishonest. $P_a$ can act dishonestly by sending the incorrect E-M1, sending the E-M3 or not sending the E-M3 at all. $P_b$ can act dishonestly by sending an incorrect signature to $P_a$ or by contacting $P_t$ before sending his signature to $P_a$. The scenarios of case 3 are discussed in cases 1 and 2 above.

- Case 4: both $P_a$ and $P_b$ are honest. If both $P_a$ and $P_b$ act honestly then fairness will be ensured in the exchange protocol and there is no need to contact $P_t$ at all.

Therefore, the above analysis of the four cases shows that the fairness is ensured either in the exchange protocol or in the dispute resolution protocol.

It is worth mentioning that $P_t$ does not need to receive any message from $P_a$ in order to resolve any dispute raised by $P_b$. Rather, $P_t$ will receive the dispute request from $P_b$ and then will decide if $P_b$'s request is valid or not. If the request is valid then $P_t$ will send the resolution electronically to both $P_b$ and $P_a$.

The certificate C-Cert is unique for each exchange. That is, every time $P_a$ and $P_b$ need to exchange their signatures on a contract then a new certificate will be used. The shared public key certificate C.$_{at}$, however, can be used for signing an unlimited number of contracts.

$P_t$ is passive during the exchange protocol i.e. in the normal execution of the protocol $P_a$ and $P_b$ will not need to contact $P_t$. In case $P_a$ misbehaves then $P_t$ will be contacted by $P_b$ to resolve the dispute.

## V. COMPARISON WITH RELATED WORK

The proposed protocol will be compared against contract signing protocols that are based on verifiable and recoverable encryption of signatures, namely, Nenadic, Zhang and Barton protocol [3], Ateniese's protocol [4] and Wang's protocol [5].

For the comparison, we analyze the number of messages and the number of modular exponentiations in both the exchange protocol and dispute resolution protocol. The exponentiation is the most expensive cryptographic operation in the finite field [5].

Both the proposed protocol and Ateniese's Protocol [4] have three messages in the exchange protocol whereas Wang Protocol [5] has seven messages. All protocols have three messages in the dispute resolution protocol.

Regarding the modular exponentiations in the exchange protocol, the proposed protocol has the lowest number of modular exponentiations, with only six. Nenadic, Zhang and Barton protocol [3] has the lowest number of modular exponentiations in the dispute resolution protocol with only five modular exponentiations. Our protocol has seven modular exponentiations in the dispute resolution protocol.

Ateniese's Protocol [4] and Wang's protocol [5] require interactive zero-knowledge proofs to allow one party to verify the encrypted signature of the other party. Our protocol offers greater efficiency in that it allows the receiving party to verify the encrypted signature using the contract certificate (C-Cert).

From Table 1, it is clear that the proposed protocol is more efficient compared with the related protocols except for the dispute resolution protocol as Nenadic, Zhang and Barton [3] protocol has the lowest number of modular exponentiations.

TABLE I.     PROTOCOLS COMPARISONS

|  | Nenadic protocol [3] | Ateniese Protocol [4] | Wang Protocol [5] | Our protocol |
|---|---|---|---|---|
| # messages in exchange protocol | 4 | 3 | 7 | 3 |
| # messages in dispute resolution protocol | 3 | 3 | 3 | 3 |
| # modular exponentiations in exchange protocol | 19 (taken from [3]) | 22 (taken from [3]) | 10.5 (taken from [5]) | 6 |
| # modular exponentiations in dispute resolution protocol | 5 (taken from [3]) | $\geq 20$ (taken from [3]) | Not mentioned | 7 |

## VI. CONCLUSION

A new offline TTP-based fair contract signing protocol is proposed in this paper. The proposed protocol ensures the exchange of signatures of two parties on a contract. At the end of the execution of the protocol, both parties get each other's signatures or neither does. The proposed protocol comprises of only three messages in the exchange protocol as well as only three messages in the dispute resolution protocol. If one party evades during the execution of the protocol, the protocol provides an online resolution for the disputes where the TTP will be involved. The proposed protocol is efficient as it has the lowest number of modular exponentiations in the exchange protocol. In a future study, we plan to investigate how to make the protocol an abuse-free protocol as Wang did in [5]. We also intend to implement and integrate the proposed protocol with e-commerce applications for the exchange of digital signatures between two parties.

AUTHOR PROFILE

**Abdullah Alaraj** is presently a faculty member in the department of Information Technology, College of Computer, Qassim University, Saudi Arabia. He received his BSc in Computer Science from King Saud University (Saudi Arabia), his MSc in Internet and Distributed Systems from Durham University(UK), and his PhD from Durham University (UK). His areas of research interests include: e-commerce security, fair exchange protocols, fraud, trust, information security